# A Survey on Unmanned Aerial Vehicle Collision Avoidance Systems

Hung Pham (1), Scott A. Smolka (1), Scott D. Stoller (1),

Dung Phan (1), Junxing Yang (1)

1. *Department of Computer Science, Stony Brook University, Stony Brook, NY, USA*

**Abstract** – Collision avoidance is a key factor in enabling the integration of unmanned aerial vehicle into real life use, whether it is in military or civil application. For a long time there have been a large number of works to address this problem; therefore a comparative summary of them would be desirable. This paper presents a survey on the major collision avoidance systems developed in up to date publications. Each collision avoidance system contains two main parts: sensing and detection, and collision avoidance. Based on their characteristics each part is divided into different categories; and those categories are explained, compared and discussed about advantages and disadvantages in this paper.

## 1 Introduction

Unmanned aerial vehicles (UAVs) have great potential to be widely used in both civil and military applications. Because of their low cost, safety benefit and mobility, UAVs can potentially replace manned aerial vehicles in many tasks as well as perform well in curriculums that tradition manned aerial vehicles do not. However, as there is no human control, UAVs usage encounters several challenges that need to be overcome; and one of those challenges is collision avoidance. In order to be used, an UAV needs the ability to surely avoid collision with both static and moving obstacles. Whereas UAV collision avoidance shares some similarities with that of air traffics and mobile robots fundamentally, UAVs surely possess many unique characteristics that need to be considered, making them an interesting research ground.

There has been a great amount of work in this field. Throughout more than two decades, many papers featuring different collision avoidance systems (CASs) have been published. Therefore a summary of them, with description, comparison and comments would be beneficial. Albaker et al. in [1] presented a nice survey of the collision avoidance approaches, summarizing their key characteristics. Alexopoulos et al. in [2] summarized some more recent collision avoidance techniques, and additionally compare between them by creating their own simulation and comparing the results. However, as more achievements are obtained in related fields (e.g. computer vision), there are more and more refined approaches to collision avoidance of the UAVs that are published recently and are not included in previous surveys.

This paper provides a survey of the major CASs from up to date related literature. We hope to give some insight and understanding of some common factors of every CAS design for UAVs, different categories of CASs as well as discussing their advantages and disadvantages.

## 2 Overview and key concepts of collision avoidance systems

The functionality of each CAS for UAV is to ensure that there is no collisions occur with some intended targets, ranging from moving and unmoving obstacles to cooperative aircrafts. In order to do that, a CAS must address the following problems:

- How to sense the environment and extract useful information about obstacles (e.g. position, speed, size, bearing angle, and so forth); and from that information, how to detect or decide that a collision is imminent.

- How the collision avoidance is performed; how the system realizes maneuvering, and possibly how to decide when to start and stop collision avoidance phase.

Based on those problems, a CAS can be divided into two main components: sensing and detection, and collision avoidance/maneuver approach. Different CASs descriptions might emphasis in different components or different sections of the components, but in general a CAS has to have both of them. Each part has its own features and design factors that can be divided into different categories, as shown in figure (1). More details on the categorization and each sub-division will be discussed in the following sections of this paper.

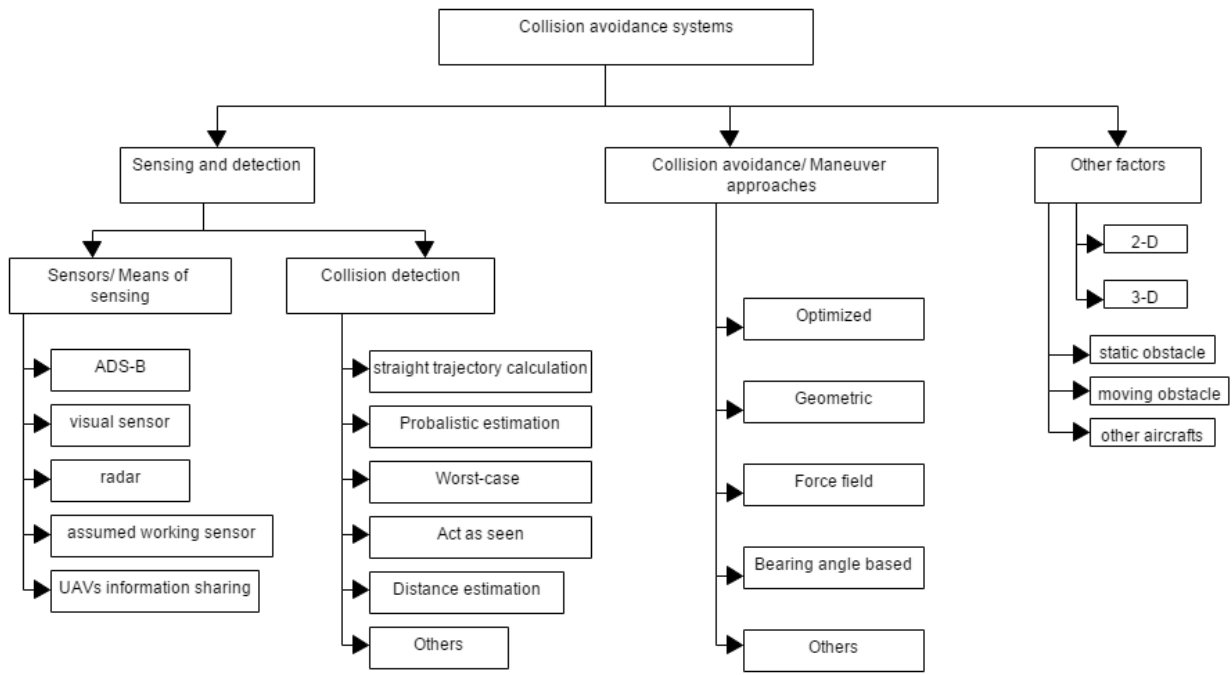

Figure (1): Illustration of CASs main factors and their sub divisions

## 3. Other design factors

Other than the 2 main components, each CAS also has certain deign factors that create the background setting of the approach. Those design factors includes sensing dimension, type of obstacles considered, type of UAV, and so forth.

The sensing dimensions of the majority of CASs are two dimensions horizontal (2-D) and three dimensions (3-D). The targeted obstacles can be static obstacles, moving obstacles or other aircrafts. The targeted obstacle type of a CAS depends greatly on the sensing mechanism, for example the other aircrafts obstacle type is mainly associated with ADS-B technology sensing method – the details will be given in the next section.

## 4. Sensing and detection

a. Sensors and means of sensing

The sensing function of a CAS is how an UAV acquires useful information about its surrounding environment. This is essential as unlike path planning, collision avoidance generally refers to the ability of the vehicle to acknowledge dangers that are not originally known and act simultaneously.

There are several types of sensor and sensing methods employed by different CASs. The most common ones are: ADS-B, visual sensor, radar, cooperative UAVs information sharing and assumed working sensor.

ADS-B, more precisely ADS-B Out, whereas ADS-B is the abbreviation for Automatic Dependent Surveillance-Broadcast, is a cooperative surveillance technology in which the aircraft periodically transmits information about altitude, airspeed, and location determined through GPS. The information is received by air traffic control ground stations and helps air traffic controllers to keep track and manage participating aircrafts. In the US, ADS-B Out equipment will become mandatory for all aircrafts in category A,B and C from 2020. For the UAVs, this technology helps to determine other friendly aircrafts in the area and to get information such as their locations, speeds and headings with accuracy. However this sensing method does not provide information of obstacles of other types. CASs that use ADS-B can be found in [3], [4], [11].

Visual sensor type usually uses cameras, for example monocular camera as in [12] and spherical camera as in (9). As those cameras return images, this sensor type often relies on image-related processes for extracting useful information about environment and obstacles. There are several advantages of visual sensor: its equipment is usually small, light, flexible and easily equipped; it returns the relative angle of the obstacles toward the UAV accurately; and in many ways this type of sensor resembles our natural vision, which makes it promising and gives researchers great intuition. The downside of this sensor type is that it restricts the information that can be reliably obtained about the obstacles, such as location, distance and size. For example, an object can appear as a small, low-contrast dot and does not change remarkably until getting very close to the vehicle. Besides, other factors such as lighting, clearness, field-of-view limit, background color, weather and so forth have very big impact on the quality of the images returned. At any rate, the approach is becoming more popular in recent literatures, partly because of achievements in the field of image processing and computer vision. This sensor type can be found in [7], [8], [9] and [12]

Radar is the typical sensor type of many transportation systems. It can scan large area quickly, are less affected by weather, dust or lighting, and can possibly have an extremely large sensing range. It can also return some specific information about the obstacles with accuracy, for example distance. The disadvantage of this sensor type is that the equipment set can be considerably big and is not suitable for UAVs. Radar appears in [10]

Cooperative information sharing occurs when in an area there are several UAVs fly with cooperation and share information about known obstacles. The information source, however, must still be collected by other sensing means. This type can be found in [6]

On the other hand, several papers do not specify the sensor type, or assume that there is a working method of sensing with some desired characteristics. We refer to it as 'assumed working sensor' type. The information returned by the assumed working sensors is usually more sufficient than that by existing sensors in order to satisfy certain requirements of the collision avoidance approach. Assumed working sensor appears in [2], [5] and [6].

b. Collision detection.

Once an UAV receives information of the obstacles, it needs to determine if there are any imminent collisions. How a CAS determines collision has a significant impact on the collision avoidance phase, for example when to start and stop collision avoidance and possibly the performance of collision avoidance. There are several collision detection approaches: trajectory calculation and distance estimation, worst case, probabilistic and act as seen.

Trajectory calculation and distance estimation are most commonly used in various CASs. Typically, collision is considered to occur if the distance between the UAV and an obstacle is less than a threshold. This approach calculates the shortest distance between the UAV and an obstacle as the UAV moves along its trajectory, or as both of them are moving in the case of non-static obstacle or another aircraft, to determine collision.

The worst case estimation appears in [4]. It considers every possibility of an obstacle trajectory and check if own UAV's trajectory intersects any of them. While being inefficient, this approach is the strongest estimation that ensures no collision can possibly happen.

Probabilistic estimation is a unique approach that is employed in [11]. In that paper a model of the UAV and the obstacle is established, with uncertain factors and their probability model distributions act as variables. More specifically, those uncertain factors are own UAV's lateral position, vertical position, along-track speed and cross-track position, and intruder UAV's heading change and altitude change. Using that model of uncertain factors, the chance that a collision happens is calculated by using Monte Carlo simulation [2]. This approach is truly interesting; however it requires remarkable computing power in order to work.

Act as seen approach, like it name suggests, does not estimate collision occurrence but rather act defensively in advance toward any obstacle the vehicle can determine and as soon as the vehicle determines it. This approach is used mainly with visual sensor. We can find this method in [5] and [9]; the method employed in [10] also shares some similarities.

*2. Monte Carlo simulation performs risk analysis by establishing models of possible results by substituting a range of values—a probability distribution—for any factor that has inherent uncertainty. It then calculates results repeatedly, each time using a different set of random values from the probability functions. Depending on the number of uncertainties and their specific ranges, a Monte Carlo simulation could involve thousands or tens of thousands of recalculations before it is complete.*

**4. Collision avoidance/maneuver approach**

With the information of obstacles returned by sensors and collision detection result as described in the previous section, a CAS must ensure that no collisions occur by applying its maneuver strategy toward imminent conflicts. This component of the CAS should not be considered a separated section – usually the collision avoidance part is closely related to the sensing and detection section, or even be a direct inference. Based on the overall characteristics of elements such as the collision avoidance approach, the maneuver trajectory, associated sensing and detection mechanism and so forth, this paper introduces five main categories of collision avoidance section: geometric, optimized trajectory, bearing angle based, force field and other types.

a. Geometric approach

This approach generally determines collisions and performs maneuver in a geometrical way, usually by simulating the trajectories of both own UAV and obstacles. In order to do that, this approach makes use of information such as location, velocity and heading of both own UAV and obstacles. Therefore, the sensing method that most approach of this type associate to is the ADS-B, making it not applicable for non-aircraft obstacles.

Park et al. in [3] demonstrates a typical example of this approach. In this publication, the method of determining collision is trajectory calculation and distance estimation. More specifically, Park calculates the subtraction of own UAV and intruder aircraft's movement vectors in a 2-D environment to determine the shortest distance between them. Then the own UAV and the intruder aircraft's trajectories are altered based on the shortest distance vector obtained, so that the shortest distance is widened in order to avoid collision. The idea is simple and straightforward; however some limitations of this approach would be needing cooperation from the intruder aircraft, and is sensitive to noises in input data from the ADS-B, which is a common factor of many collision avoidance approaches that require exact calculation.

Another instance of this approach can be found in [4] by Strobel et al. This paper employs the worst case collision detection in 2-D environment: assuming the intruder aircraft maximum turn rate, one can establish a threat region where the aircraft can possibly be in a short time, e.g. 30 seconds, in the future. Collision threat is detected if own UAV position after that time is inside the threat region. For maneuvering, one new heading is generated as the old heading plus or minus 90 degrees depending on the heading of the intruding aircraft; and after the threat is gone the UAV returns to original trajectory. This approach can possibly relieve the weakness of sensitivity to input noise by employing the strongest collision detection mechanism.

b.  Optimized trajectory approach

This type of approach shares some similarities with the previous one, as it relies on some trajectory calculation in a geometrical way. However this approach possesses one distinct characteristic: the trajectory generated usually is the most optimized one –the UAV can ensure to avoid all the obstacles whereas still maintain a good closeness with the predetermined trajectory toward its desired targets.

This approach type shares a great similarity with the path planning problem of an UAV, which is finding an efficient trajectory toward desired targets while avoiding pre-known obstacles. The main obstacles targeted by this approach are static. Besides, in order to generate a good trajectory, the UAV must sufficiently collect information about obstacles such as position and size. Consequently, the usual sensing method employed in CASs of this type is assumed working sensor, making it more theoretical than practical. On the other hand, similar to many optimization processes, the calculation amount required further limits its practicality, considering the limited processing power of an UAV and the limited time to act before collision. Regardless, the trajectory generated contains some desirable attributes; and this approach type has good studying value for researchers.

Two important examples of this type can be found in [2] and [6]. Boivin et al. in [6] describe a model in 3-D to represent the UAV, taking in parameter of time and can predict future coordinates of the UAV by considering the possible commands it will take in a short time. From the UAV current position and the destinations' coordinates, a cost function is formulate such that minimizing it results in the best set of future commands and the optimal trajectory. For each of that best set of future commands, a method similar to trajectory calculation and distance estimation is applied for collision detection. If this constraint fails, another set of 'close to best' future commands is evaluated; the system might have to recalculate the cost function several times in the process.

Another method is mentioned by Alexopoulos in [2]. Assumed the obstacles position, size and shape are known in advance, this approach divides the remaining 2-D map in a grid and represent them as a weighted graph. A collision free path, while still maintaining good closeness with the UAV original trajectory, can be found with the help of graph search algorithms such as A*algorithm (3).

Both of these two CASs have the same weakness of the collision avoidance type.

*3. A\* uses a best-first search and finds a least-cost path from a given initial node to one goal node (out of one or more possible goals). As A\* traverses the graph, it follows a path of the lowest expected total cost or distance, keeping a sorted priority queue of alternate path segments along the way.*

c.  Bearing angle based approach

This approach is an interesting, relatively new approach that utilizes the use of visual sensor and its ability to accurately return the relative angle of the obstacles toward the UAV. The main idea is that by keeping obstacles' images at a 'safe' position in the sensor field of view, the UAV can effectively prevent collision. This type of approach features spiral flight paths, as it has been proved that 'the path of an aircraft flying at a constant velocity and with a constant relative bearing to a stationary object constructs an equiangular spiral trajectory' by Yang in [8].The bearing angle approach is affected by all the disadvantages of the visual sensor, which are relying on image features processing techniques and significantly affected by external conditions.

[7], [8] and [9] are all examples of this approach. Whereas having the similar general idea, the details in those papers are different in many ways. [7] has a sensing dimension of 2-D, while the dimensions of [8] and [9] are 3-D. In [7], Saunders used the camera's focal length to estimate the distance of the UAV toward the obstacle. This is not very clear, however, since an obstacle's image on the image plane also depends on how big the obstacle is.  In [8] Yang collects different samples of the heading angles as the UAV moves in its spiral trajectory toward the obstacle in order to determine the relative distance. In both papers, that distance is used as the trigger for collision avoidance start and stop phase. In [9], McFayden does not compute the distance and completely relies on the obstacle bearing and the vehicle heading to perform collision avoidance. The paper utilizes act as seen method: starts maneuvering as soon as the UAV detects an obstacles; and for stopping collision avoidance phase, it uses an objective function that minimizing results as the vehicle returns to its initial heading.

Even though visual sensor has a lot of limitations, with the improvements in image processing and computer vision as well as various studies and researches, the use of this sensor type, and in particular the bearing angle approach, is becoming more frequent.

d. Force field approach

The force field approach has appeared for a long time, and was used rather frequently in papers one decade ago. With the intuition taken from the repulsive and attractive electrical force field, using the concept of 'potential' as a measurement of desirable characteristics of the UAV trajectory, the approach can calculate a good trajectory that is collision free and satisfy other requirements related to destination, path complexity, vehicle velocity and so forth. Apparently despite some differences in details and in the trajectory generated, this approach shares a large number of similarities with the optimized approach; the reason why this paper considers it a separated category is because the force field was a major approach in the past publications.

Miura et al. in [5] present a typical example of this approach. The paper uses assume working sensor and act as seen collision detection method, with obstacle type is other aircrafts. Five desired conditions were chosen for estimating the 'potential': the distance toward other airplanes, closeness to the original path, small velocity change, and the lack of sudden maneuvers and complex maneuvers. If the potential is small, the vehicle is in a desirable route and vice versa. However, calculating potential of each point on the plane is a huge work; therefore the gradient of the potential is used instead. As the present and future location of the vehicles and intruder aircrafts are computed, the potential gradient at each sampling point is calculated. The process is completed if the maximum value of potential gradient is less than a threshold; otherwise the sampling points are shifted and the process is repeated. Note that it may take thousands of repetitions in order to achieve a good solution.

Similar to the optimized approach, the force field method requires remarkable calculating power and time, which make it unsuitable for UAVs real life application.

e. Other types

Viquerat et al. in [10] demonstrates a CAS that uses radar sensor. The radar is broadcasted every timeframe, and the UAV chooses the trajectory according to the region with the lowest return radar signal. This process is repeated every 0.1 second, which is applicable because of the stateless machine associates with the radar can perform better with rapid analysis of time-variant data than the normal state-ful machine. However, the UAV needs to keep track of its heading and position every time it performs maneuver to determine when it returns original track; and in the case a continuous collision-free path does not exist, the UAV will be lost.

Kim et al. in [11] presents a CAS that focuses on the probabilistic collision detection method. The sensor type is ADS-B and obstacle type is other aircrafts; and the maneuver approach is pre-defined according to different probability of collision. The idea of calculating probability of collision is natural and inresting approach; however it requires huge processing capacity in order to simulate all the Monte Carlo simulations, and the approach does not ensure collision-free trajectory.

Saha et al. in [12] presents an interesting approach based on visual sensor and geometrical calculation in 3-D. In the paper, the CAS uses two monocular cameras. By using descriptor vectors of the two images and the marching between them, the obstacle (which is a group of feature points) can be detected; this is based on a computer vision technique name SURF. For every feature point, from its two images by the two cameras and the cameras' focal lengths, a system of equations can be established to finally calculate the geometric location of that point. The heading of the UAV is changed if the distance of any feature point to the UAV is less than a threshold. This is a truly promising approach; with more careful implementation on the maneuver trajectory it can achieve significant practical usefulness.

**5. Conclusion**

In this paper, we have presented a survey of major collision avoidance systems from various papers from the past to the recent. We have discussed aspects of different components of the CASs, as well as the advantages and disadvantages of the approaches to different problems in collision avoidance.

*This is a draft. In the near future, this paper needs to be updated with figures and with the more recent papers that were found but not yet included, as well as looking for a more suitable way of categorizing the CASs if possible.*

**6. References**

1. Albaker, B.M.; Rahim, N.A., "A survey of collision avoidance approaches for unmanned aerial vehicles," in *Technical Postgraduates (TECHPOS), 2009 International Conference for* , vol., no., pp.1-7, 14-15 Dec. 2009

2. Alexopoulos, A.; Kandil, A.; Orzechowski, P.; Badreddin, E., "A Comparative Study of Collision Avoidance Techniques for Unmanned Aerial Vehicles," in *Systems, Man, and Cybernetics (SMC), 2013 IEEE International Conference on* , vol., no., pp.1969-1974, 13-16 Oct. 2013

3. Jung-Woo Park; Hyon-Dong Oh; Min-Jea Tahk, "UAV collision avoidance based on geometric approach," in *SICE Annual Conference, 2008* , vol., no., pp.2122-2126, 20-22 Aug. 2008

4. Strobel, A.; Schwarzbach, M., "Cooperative sense and avoid: Implementation in simulation and real world for small unmanned aerial vehicles," in *Unmanned Aircraft Systems (ICUAS), 2014 International Conference on* , vol., no., pp.1253-1258, 27-30 May 2014

5. A. Miura, H. Morikawa, M. Mizumachi, "Aircraft Collision Avoidance with Potential Gradient Ground Based Avoidance for Horizontal Meneuvers", Electronics and Communications in Japan, Part 3, vol. 78, no. 10, pp. 104-113, 1995

6. Boivin, E.; Desbiens, A.; Gagnon, E., "UAV collision avoidance using cooperative predictive control" in *Control and Automation, 2008 16th Mediterranean Conference on* , vol., no., pp.682-688, 25-27 June 2008

7. Saunders, J., Beard, R., "Reactive vision based obstacle avoidance with camera field of view constraints", in AIAA Guidance, Navigation and Control Conference and Exhibit. Honolulu, Hawaii (2008)

8. X. Yang, L. Mejias, and T. Bruggemann, "A 3D collision avoidance strategy for uavs in an non-cooperative environment," in Journal of Intelligent Robotic Systems, vol. 70, No. 1-4, pp. 315-327, April 2013.

9. McFadyen, A.; Durand-Petiteville, A.; Mejias, L., "Decision strategies for automated visual collision avoidance," in *Unmanned Aircraft Systems (ICUAS), 2014 International Conference on* , vol., no., pp.715-725, 27-30 May 2014

10. Viquerat, A., Blackhall, L., Reid, A., Sukkarieh, S., Brooker, G., "Reactive collision avoidance for unmanned aerial vehicle using Doppler radar", in: 6th International Conference on Field and Service RoboticsFSR 2007 (2007)

11. Kwang-Yeon Kim; Jung-Woo Park; Min-jea Tahk, "UAV collision avoidance using probabilistic method in 3-D," in *Control, Automation and Systems, 2007. ICCAS '07. International Conference on* , vol., no., pp.826-829, 17-20 Oct. 2007

12. Saha, S.; Natraj, A.; Waharte, S., "A real-time monocular vision-based frontal obstacle detection and avoidance for low cost UAVs in GPS denied environment," in *Aerospace Electronics and Remote Sensing Technology (ICARES), 2014 IEEE International Conference on* , vol., no., pp.189-195, 13-14 Nov. 2014

13. Mejias, L.; McNamara, S.; Lai, J.; Ford, J., "Vision-based detection and tracking of aerial targets for UAV collision avoidance," in *Intelligent Robots and Systems (IROS), 2010 IEEE/RSJ International Conference on* , vol., no., pp.87-92, 18-22 Oct. 2010

14. Ramasamy, S.; Sabatini, R., "A unified approach to cooperative and non-cooperative Sense-and-Avoid," in *Unmanned Aircraft Systems (ICUAS), 2015 International Conference on* , vol., no., pp.765-773, 9-12 June 2015

15. McFadyen, A.; Mejias, L., "Design and evaluation of decision and control strategies for autonomous vision-based see and avoid systems," in *Unmanned Aircraft Systems (ICUAS), 2015 International Conference on* , vol., no., pp.607-616, 9-12 June 2015

16. McFadyen, A.; Mejias, L.; Corke, P.; Pradalier, C., "Aircraft collision avoidance using spherical visual predictive control and single point features," in *Intelligent Robots and Systems (IROS), 2013 IEEE/RSJ International Conference on* , vol., no., pp.50-56, 3-7 Nov. 2013